# Dynamical Phenomena in an Optical-Wavelength Phonon Laser (Phaser): Nonlinear Resonances and Self-Organized Mode Alternation

D. N. Makovetskii

(*Institute for Radio-Physics and Electronics NASU, Kharkov, Ukraine*)

**Abstract**

This is a part of an overview of my early studies on nonlinear spin-phonon dynamics in solid state optical-wavelength phonon lasers (phasers) started in 1984. The main goal of this work is a short description and a qualitative analysis of experimental data on low-frequency nonlinear resonances revealed in a nonautonomous ruby phaser. Under phaser pumping modulation near these resonances, an unusual kind of self-organized motions in the ruby spin-phonon system was observed by me in 1984 for the first time. The original technique of optical-wavelength microwave-frequency acoustic stimulated emission (SE) detection and microwave-frequency power spectra (MFPS) analysis was used in these experiments (description of the technique see: D.N.Makovetskii, Cand. Sci. Diss., Kharkov, 1983). The real time evolution of MFPS was studied using this technique at scales up to several hours. The phenomenon of the self-organized periodic alternation of SE phonon modes was experimentally revealed at hyperlow frequencies from about 0.1 Hz to 0.001 Hz and less (for the phaser pumping modulation at approximately 10 Hz). The nature of this phenomenon was investigated by me later in details (see: arXiv:cond-mat/0303188v1 ; arXiv:cond-mat/0410460v1 ; Tech. Phys. Letters, 2001, Vol.27, No.6, P.511-514 ; Tech. Phys., 2004, Vol.49, No.2, P.224-231).

Stimulated emission (SE) of optical-wavelength microwave-frequency acoustic waves was detected and studied as early as 1960's – 1970's in dielectric crystals doped with paramagnetic ions under conditions of inversion of their spin-levels populations [1 - 8]. Due to the quantum nature of this laser-like phenomenon, the emitted acoustic flow may be interpreted as an ensemble of quasiparticles – phonons. So, by an analogy with the term *laser* (light amplification by stimulated emission of radiation), the term *phaser* (phonon amplification by stimulated emission of radiation) was introduced by U.Kh.Kopvillem for devices based on the phonon SE phenomenon [9].[1] We will use here this last term.

---

[1] There are some other terms (Phonon Maser, Saser, Uaser etc)., which are applied by various authors both to SE-systems of this class and to another laser-like systems with acoustic SE (e.g., suboptical-wavelength optical-frequency acoustic lasers).

## Autonomous phaser

In an autonomous phaser, the acoustic SE leads to a stationary amplification [1, 3, 4, 6] of an injected optical-wavelength sound, called also microwave ultrasound or *hypersound*. This simplest experimentally detectable phenomenon takes place in the subthreshold regime of a phaser, when total amplification $\alpha$ of an injected hypersound is less by modulo than its total dissipation $\alpha_d$, i.e. when $\alpha + \alpha_d > 0$.

In the superthreshold regime of an autonomous phaser, the last inequality is reversed, and the phenomenon of the SE causes an instability (self-excitation) of optical-wavelength acoustic waves. A dissipative crystalline system with well-resolved microwave acoustic resonances, e.g. acoustical Fabry-Perot resonator (AFPR), works in this case as mode selector. As the result, the acoustic SE in a crystalline AFPR leads to single-mode or (more typically) multimode generation of phonons, which is an analog of generation of photons in usual electromagnetic lasers.

These two basic regimes (subtreshold amplification and superthreshold generation) are very typical for autonomous phasers as well for autonomous lasers. More complicated an sometimes unexpected regimes for autonomous SE-systems (self-induced collapse of amplification, subthresold generation etc.) are possibly too. These phenomena will be considered in other parts of our work. Here we will concentrate our attention on unusual dynamics of a *nonautonomous* phaser.

## Nonautonomous phaser

A nonautonomous dynamical dissipative system has at least one time-dependent control parameter. Formally, periodically modulated (nonautonomous) dynamical system may be considered as an autonomous one, but in phase space of a higher dimension. So, the simplest 2D Tang-Statz-deMars nonlinear autonomous laser model [10] must be expanded to a 3D model [11], if a single control parameter is periodically modulated.

Third dimension of a dissipative system permits complex nonperiodic motions in the phase space. Moreover, hypersensitivity to initial conditions in many typical laser and laser-like 3D models is the cause of long-time inpredictability of formally deterministic motion (low-dimensional deterministic chaos). These phenomena were observed experimentally by us in ruby phaser modulated at frequency of nonlinear $\nu$-resonance, which is the acoustical analog of the laser nonlinear relaxational resonance [11].

In this work, we report on the first experimental observation of the series of qualitative new nonlinear resonances in a nonautonomous phaser, which, to our knowledge, were not observed in usual, electromagnetic lasers. We call them $\lambda$-resonances. These new resonances were observed by us at frequencies ranges situated beyound of the $\nu$-resonance domain. All resonances of this series are very narrow, and the fundamental frequency $\omega_\lambda$ of the series is about 9.8 Hz, i.e. it is of order smaller than the relaxational frequency $\omega_\nu \approx 10^2$ Hz in our ruby phaser.

## Conditions of the experiments

Experiments were carried out at liquid-helium temperatures (below 2 K) on artificial pink ruby crystal doped with three-valent chromium ions (at atomic concentration 0.03%). The rod of 2.6 mm diameter and 17.6 mm length with optically plane and parallel faces (acoustical mirrors) was produced from this crystal. It served as AFPR with acoustical quality $Q_A \approx 10^6$ for longitudinal phonon modes at frequencies about $9 \cdot 10^9$ Hz.

The piesoelectric transducer has the form of thin half-acoustic-wavelength textured ZnO film with Al sublayer. The transducer was used both for detection of an optical-wavelength acoustic SE in the AFPR and (optionally) for injection of non-SE optical-wavelength acoustic signal from outside.

The inversion states of chromium active centers were formed by the electromagnetic field pumping at the frequency about $2.3 \cdot 10^{10}$ Hz by means of the cylindrical cavity of the $H_{011}$ type with quality factor $Q_P \approx 10^4$. The AFPR was placed inside of the pumping cavity along its geometrical axis. A modulated klystron generator was used as a source of pumping.

The microwave heterodyne spectrum analyser with the frequency resolution about 1 kHz was used for observation of phaser acoustic power spectra (using electromagnetic signal linearly converted from acoustic one by the transducer).

All the measurements were carried out under direction of static magnetic field H at an angle $\vartheta_{symm} = \arccos(1/\sqrt{3})$ to the crystallographic axis $\vec{C}$ of ruby coinsiding with the geometrical axis of the AFPR when the push-pull pumping conditions are realized.

## Results of experiments

The real time evolution of MFPS was studied at scales up to several hours. At pumping modulation near $\omega_m = 10$ Hz (and at $2\omega_m$, $3\omega_m$) an unusual kind of self-organized motions in the phaser spin-phonon system was observed for the first time. The general picture of the phenomenon revealed is as follows.

Evolution of acoustic SE power spectra of modulated phaser is laminar and has the form of periodic and very slow (comparing to $\omega_m^{-1}$) mode alternations. There are sequential ignition of new and new SE modes at one side of the microwave power spectrum and extinguishing approximately of that SE mode number on the other side of the microwave power spectrum. This mode alternation process takes place up to the full break-down of motive SE spectrum at the determined "finish" frequency domain. Then a new group of SE modes are ignited at determined "start" frequency domain and so on.

Generally, the acoustic SE power spectrum evolution demonstrate regular sequential motions of emission *spectral area as a whole* but under keepeng of localization of *each individual SE mode* at positions of the AFPR modes.

The period of full cycle of such self-detunings $\tau_p^{(\lambda)}$ is changing by several orders when $\omega_m$ is scanning in a narrow range (less than 1 Hz) in the vicinity of the revealed nonlinear $\lambda$-resonance, $\omega_\lambda \approx 9.8$ Hz.

Further experiments showed that the sign of a derivative $d\Omega_B/dt$ (where $\Omega_B$ is the biggest-amplitude frequency of the motive spectrum section) is determined by the sign of the **detuning** $\Delta_\lambda$ of the modulaton frequency $\omega_m$ from the central frequency $\omega_\lambda$ of the $\lambda$-resonance such as $\text{sgn}(d\Omega_B/dt) = -\text{sgn}\,\Delta_\lambda$, where $\Delta_\lambda = \omega_m - \omega_\lambda$.

The time dependencies $J_\sigma(t)$ of the integral SE intensity $J_\sigma$ in the $\lambda$-resonance domain were periodic – in the contrary to the $\nu$-resonance. By approaching $|\Delta_\lambda|$ to zero the full cycle period $\tau_p^{(\lambda)}$ of the SE self-detunings increases to values which are of 4 – 5 orders greater than $\omega_m^{-1}$.

So, in the $\lambda$-resonance domains, a large-scale spatially-temporal ordering takes place for our phaser active system under periodic external forcing. This is in contrast to the $\nu$-resonance domain [12, 13], where increasing of the external force destroys intermode correlations and leads to deterministic chaos.

On the other hand, the spin-phonon ordering in the $\lambda$-resonance domain is essentially nontrivial. This ordering (periodic alternation of phonon SE modes) does not reproduce periodicity of the external force. The period of full cycle of such alternations is generally many orders greater than period of the external force $\tau_m \equiv 2\pi/\omega_m$. Moreover, small (some percent) changing of $\tau_m$ leads to huge changing of $\tau_p^{(\lambda)}$ in all experiments $\lambda$- resonance domain. Such induced by an external force ordering, which is uncorrelated with the external force period, talks about self-organized nature of the phenomenon observed. Naturally, a question about quantitative analysis of the experimental data arises. A first, but very important step in this direction is made in the following Section.

## Why the $\lambda$-resonance is at this frequency?

Let us discuss a possible mechanism of $\lambda$- resonance formation based on the classical balance SE model [10], taking in account the papers [14, 15] on the nonlinear dynamics in multimode lasers. In our work it will be studied the simplest two-mode model with homogeneous transverse SE field. Following to [15] we consider the 5D autonomous flow system and introduce the 1D external periodic perturbation. The 5D vectorial order parameter **B** for the two-mode phaser reads:

$$\mathbf{B}(t) = (J_1, J_2, D_0, D_1, D_2), \qquad (1)$$

where $J_i(t)$ – normalized dimensionless phonon SE intensity for the $i-$th mode ($i=1;2$); $D_j(t)$ – Fourier components of the inverse spin-level population difference in active medium ($j=0;1;2$):

$$D_j(t) = \frac{1}{L_C} \int_0^{L_C} D(x,t) \cos(2k_j x)\, dx. \qquad (2)$$

Here $D(x,t)$ – spatially-temporal distribution of inverse spin-level population difference normalized to its thermodynamically equilibrium value; $k_0 = 0$; $k_{1,2}$ – wavenumbers for 1-th and 2-th mode respectively, the numbering ($s=1,2$) follows the order of the diminishing of their phaser amplification factor $\alpha_s$, i.e. $\mu_1 = 1$; $\mu_2 = \alpha_2/\alpha_1 < 1$, where $\mu_s = \alpha_s/\alpha^{(m)}$; $\alpha^{(m)} \equiv \max(\alpha_s) = \alpha_1$; $L_C$ – AFRP length ($L_C \gg k_{1,2}^{-1}$).

Let us use the adiabatic approximation, i.e. let $\tau_2 \ll \tau_1, \tau_C$, where $\tau_1$ and $\tau_2$ are the longitudinal and the transverse relaxation times of the active centers, $\tau_C$ is the lifetime of resonant phonons in AFRP. Following [15], we can reduce spatially-1D task on two-mode phonon SE to the flow equations with spatial Fourier components of spin populations. These equations, in countrary to the standard Tang-Statz-deMars model [10], describes our nonequilibrium dissipative system *by the account of the intermode interaction* under conditions of weak spin diffusion:

$$\tau_1 \frac{d\mathbf{B}}{dt} = \vec{\psi}^{(L)}(\mathbf{B}) + \vec{\Psi}^{(NL)}(\mathbf{B}), \qquad (3)$$

where $\vec{\psi}^{(L)}$ and $\vec{\Psi}^{(NL)}$ – the linear and nonlinear terms of the right-hand part of the equation, having such the vectorial components:

$$\left.\begin{array}{l} \psi^{(L)}_{1,2} = -2 J_{1,2}/q_1; \quad \psi^{(L)}_3 = A - D_0; \\[4pt] \psi^{(L)}_{4,5} = -D_{1,2}; \\[4pt] \Psi^{(NL)}_{1,2} = (2D_0 - D_{1,2})\mu_{1,2} J_{1,2}/q_1 \\[4pt] \Psi^{(NL)}_3 = [(D_1 - 2D_0)\mu_1 J_1 + (D_2 - 2D_0)\mu_2 J_2]/2; \\[4pt] \Psi^{(NL)}_{4,5} = \mu_{1,2} D_0 J_{1,2} - (\mu_1 J_1 + \mu_2 J_2) D_{1,2} \end{array}\right\}, \quad (4)$$

where $A$ is pumping parameter; $q_1 \equiv 2\tau_C/\tau_1 \equiv 2B$.

Due to $q_1 \ll 1$ (it is $q_1 \approx 10^{-4} \div 10^{-5}$ in our experiments), and using results of work [15], we find that besides of nonlinear resonance at $\omega_v$, which arises even in the supermode approximation, it is once more resonance, caused by the intermode energy exchange. This resonance appears from the complex-conjugated Liapunov exponents $\Lambda_{3,4}$ for the stationary states of the flow (3). The imaginary part of $\Lambda_{3,4}$ (omitting the sign) reads as follows:

$$\mathrm{Im}(\Lambda_{3,4}) \approx \left(\frac{J_2^{(\mathrm{st})}}{4\tau_C}\right) \frac{[4(1+\mu_2) - 3\mu_2 D_0^{(\mathrm{st})}]\mu_2 D_0^{(\mathrm{st})} - 4(1-\mu_2)^2}{4 - D_0^{(\mathrm{st})}} \ll \omega_\nu, \quad (5)$$

where index $^{(\mathrm{st})}$ labels the stationary solutions of (3) in the two-mode phonon SE domain:

$$A > A_{\mathrm{two}} \equiv \frac{4\mu_2 - 2\mu_2^2 - 1}{(2\mu_2 - 1)\mu_2}. \quad (6)$$

In that way the interaction of the phaser modes brings up additional characteristic frequency of the system, which may be identified to the experimentally revealed nonlinear resonance frequency $\omega_\lambda \ll \omega_\nu$. Moreover, the numerical solutions of the full system of laser equations, analogous to (3) with right pars of the form (4), at $A_0 = 1,8$; $\mu_2 = 0,83$; $q_1 = 10^{-4}$ (i.e. not far from our experimental conditions), demonstrated so called antiphase motions for two-mode SE. In $N$-mode systems antiphase motions are of more complicated type (see e.g. [14]), but it is preserved their common nontrivial feature – coherent unidirectional pulsations of the SE modes with the temporal delay of $\tau_p^{(\lambda)}/N$ for the nearest neighbors. Accordingly, we interpret revealed phenomenon of phonon spectral autowave-like motions as the result of arising of antiphase states for SE in AFPR under destabilization of stationary mode spatial distribution by the external force at $\omega_m \approx \omega_\lambda$.

### Final notes

Experimental results in this field were primarily published by me in 1989 in Russian [16]. The nature of the phenomenon described above was experimentally investigated later in details – see [17 - 21]. Theoretical interpretation of the self-organized alternation of phonon SE modes (based on antiphase dynamics [14, 15]) was proposed by me in [17, 19, 22]. Some early attempts to understand the nature of the observed mode alternation were published in [23, 24]. The original experimental technique (optical-wavelength microwave-frequency acoustic SE detection and the phonon MFPS analysis) used in these experiments was partially described in my dissertation on phaser dynamics [25, 26]. More details on the used experimental technique see [19, 21]. An alternative theoretical approach concerning phenomena of self-organization in phasers and some other excitable systems was developed in [27 - 34].

## List of abbreviations

AFPR – Acoustical Fabry-Perot Resonator

MFPS – Microwave-Frequency Power Spectra

PHASER (phaser)– Phonon Amplification by Stimulated Emission of Radiation

SE – Stimulated Emission